\documentstyle[twoside,fleqn,espcrc2]{article}

\newcommand{\AmS}{{\protect\the\textfont2
  A\kern-.1667em\lower.5ex\hbox{M}\kern-.125emS}}

\title{Charged Pion Polarizability From the Lattice}

\author{Walter Wilcox\address{Department of Physics, Baylor University \\
 Waco, TX 76798-7316}}%

\begin{document}

\begin{abstract}
Direct evaluation of charged particle polarizabilities on the lattice
is quite difficult. However, a 
short cut for charged pion polarizability - the Das, Mathur,
Okubo Sum Rule - can readily be calculated using lattice techniques. 
A phenomenological model has been developed to fit the time behavior of 
the propagators in this expression. 
Numerical systematics are discussed and some preliminary results 
are presented.
\end{abstract}

\maketitle

\section{Introduction}

External field methods have been used previously to measure
neutral particle polarizabilities in the context of lattice QCD\cite{fiebig}.
However, there have been no investigations of charged hadron polarizabilities
using the lattice techniques. This is unfortunate since
there is now a growing body of experimental information on
this subject. Of course, external field methods would be very difficult or
impossible to use in the case of charged hadrons because charged particles 
accelerate in an electric field. The evaluation of charged 
particle polarizabilities can be done on the lattice by measuring
a Compton scattering coefficient\cite{II}. However,
the direct evaluation is actually quite difficult because of
the many diagrams involved.

There is a way around this problem if one is willing to
work in the exact chiral limit. Let us examine a sum rule
for pion polarizability and see how readily it can be 
evaluated on the lattice. Conventional wisdom is that 
charged pion polarizability involves cancellation of
large numbers, so may be difficult to simulate.

\section{DMO Sum Rule and Model}

The Das, Mathur, Okubo Sum rule for the charged pion polarizability, derived
from current algebra in the chiral limit is\cite{first}:
\begin{eqnarray}
\lefteqn{\alpha_{E}= \alpha \frac{<r^{2}>}{3m_{\pi}} } \nonumber \\
& &+\frac{\alpha}{m_{\pi}F_{\pi}^{2}}
\int^{\infty}_{4m_{\pi}^{2}}\frac{ds^{2}}{s^{4}}(\rho_{A}(s^{2})
 -\rho_{V}(s^{2})).
\label{DMO}
\end{eqnarray}
This employs only mesonic two-point functions, which are
easily calculable on the lattice. The spectral integrals
can be measured using a combination of numerical techniques, actually 
developed other physics contexts (the \lq\lq volume method"\cite{vol} 
and \lq\lq Fourier reinforcement"\cite{four}), which 
results in a very good signal. The value of the term proportional to the
pion squared charge radius, $<r^{2}>$, and the pion decay constant, $F_{\pi}$, are
taken from experiment. One can change the spectral integral above into a
lattice space-time integral through the use of (the momentum transfer is 
purely spatial, $q_{\mu} =(0,\vec{q})$)
\begin{eqnarray}
\lefteqn{\frac{d}{d\vec{q}^{\,2}}
(\Delta^{V}_{00}(\vec{q}^{\,2})-\tilde{\Delta}^{A}_{00}
(\vec{q}^{\,2}))|_{\vec{q}^{\,2}=0}=} \nonumber \\
& & \quad\quad\quad\quad\int^{\infty}_{4m_{\pi}^{2}}\frac{ds^{2}}{s^{4}}(\rho_{A}(s^{2}) -
\rho_{V}(s^{2})),
\label{this}
\end{eqnarray}
where in Euclidean infinite space ($V_{0}(x)\equiv \bar{d}(x)\gamma_{0}u(x)$)
\begin{eqnarray}
\lefteqn{\Delta^{V}_{00}(\vec{q}^{\,2})=} \nonumber \\
& & \int d^{4}x e^{-iq\cdot x}<0|T(V_{0}(x)V_{0}^{\dagger}(0))|0>.\label{that}
\end{eqnarray}
(The subtracted axial propagator, $\tilde{\Delta}^{A}_{00}(\vec{q}^{\,2})$, is
similar to Eq.(\ref{that}) but has the pion pole removed.) The left-hand side of
Eq.(\ref{this}) is formed by taking a numerical momentum derivative of the lattice
data. 

There will be new sources of systematic uncertainty in the lattice evaluation
of sum rules. In this case the above shows that one must perform a numerical
momentum derivative. In addition, the time integral can be performed
in either a discrete or continuous sense, with attendant uncertainty. 
One can simply apply Simpson's rule in the discrete case. 
The continuous case demands some way of interpolating between the 
lattice time points. For this purpose, let us consider the standard lattice
point-to-point charge density correlator:
$$
\sum_{\vec{x}}e^{-i\vec{p}\cdot \vec{x}}<0|T(V_{0}(x)V_{0}^{\dagger}(0))|0>.
$$
One may show that (all masses and distances dimensionless) this reduces to
$$
\sum_{\vec{x}}e^{-i\vec{p}\cdot
\vec{x}}Tr[S(x,0)\gamma_{4}\gamma_{5} S^{\dagger}(x,0)\gamma_{4}\gamma_{5}],
$$
where $S(x,y)$ is the quark propagator and the trace is over color and Dirac
spaces. Use the free quark propagator 
(coordinate gauge; diagonal to this order in color space):
\begin{equation}
S(x,0) = \frac{1}{2\pi^{2}}\frac{\gamma\cdot x}{x^{4}}+\frac{1}{(2\pi)^{2}}
\frac{m_{q}}{x^{2}}+ \cdots.
\end{equation}
Make the replacement $\sum_{x}\rightarrow \int d^{3}x$ and define 
\begin{equation}
G^{V}_{00}(t,\vec{p})\equiv \int d^{3}x e^{-i\vec{p}\cdot
\vec{x}}\frac{t^{2}-r^{2}}{(r^{2}+t^{2})^{4}}.
\end{equation}
The next term is of order $m_{q}^{2}$ and is negligible in our phenomenological
fits. Do the angular integrals ($p \equiv |\vec{p}|$):
\begin{eqnarray}
\lefteqn{G^{V}_{00}(t,\vec{p}) =} \nonumber \\
\lefteqn{\frac{12}{\pi^{3}p}\int_{0}^{\infty} dr r \sin
(pr) \{\frac{2t^{2}}{(r^{2}+t^{2})^{4}}-
\frac{1}{(r^{2}+t^{2})^{3}} \} .}
\end{eqnarray}
Actually, what one wants in this case is
the derivative of the above with respect to $p^{2}$ (see Eq.(\ref{this}) above).
I prefer to form the same numerical derivative in the phenomenological model as
is necessary on the lattice. Thus, in order to compare with the lattice data,
consider
\begin{eqnarray}
\lefteqn{{\cal G}^{V}_{00}(t,\vec{p})\equiv
\frac{G^{V}_{00}(t,\vec{p})-G^{V}_{00}(t,0)}{p^{2}}} \nonumber \\
& &=\frac{12}{\pi^{3}p^{2}}\int_{0}^{\infty} dr r [\frac{\sin
(pr)}{p}-r] \nonumber \\
& &\times\{\frac{2t^{2}}{(r^{2}+t^{2})^{4}}-
\frac{1}{(r^{2}+t^{2})^{3}} \}.
\end{eqnarray}
To this order the same functional form holds for the axial propagator as
well. 

There are two modifications to this function that are 
necessary before one compares to the lattice data. First, it is clear 
that the above expressions have an ultraviolet infinity associated 
with the $r=0$ lower limit. This infinity can be controlled, 
as the lattice itself controls it,
by putting in a short distance cutoff. So, replace the lower limit
above by $r_{0}>0$, which becomes a parameter in the fits. Call
this modified function ${\cal G}^{V}_{00}(t,\vec{p},r_{0})$. Second, put in a
continuum threshold, $s_{0}$, to control the onset of excited states in the
spectral density. One can easily show that the resulting function is now
given by ($t>0$ understood)
\begin{eqnarray}
\lefteqn{\hat{{\cal G}}^{V}_{00}(t,p,r_{0},s_{0})\equiv} \nonumber \\
& &\zeta\int_{s_{0}}^{\infty}ds\int_{-\infty}^{\infty}
\frac{du}{2\pi}e^{ius}{\cal G}^{V}_{00}(t+iu,\vec{p},r_{0}).\label{theother}
\end{eqnarray}
A third parameter, $\zeta$, has been added in Eq.(\ref{theother}) 
in order to account for lattice anisotropy at small lattice times as in
Ref.\cite{lein}. One can analytically do the two integrals in 
Eq.(\ref{theother}); 
the remaining oscillating radial integral is done
numerically. One must also apply vector and axial vector
renormalization constants to the lattice data before comparing to
Eq.(\ref{theother}). In addition, add to Eq.(\ref{theother}) a function
$$
\lambda^{2}e^{-m_{V,A}t},
$$
to fit the lowest pole mass in the vector (rho meson) or axial vector
(A1 meson) cases. These mass values are fixed from previous independent lattice
measurements, so all told there are 4 parameters in the model:
$\zeta$, $\lambda$, $s_{0}$ and $r_{0}$.

Using Wilson fermions at $\beta=6.0$ on a $16^{3}\times 24$ lattice, one has
$p_{min}=\frac{\pi}{8}$, and this value is used in both the numerical lattice
derivative as well as the phenomenological model. I have not yet done the
chiral extrapolation in quark mass, but the results from Eq.(\ref{DMO}) when 
plotted as a function of quark mass are extremely flat.

The fit vector correlator (unnormalized) is shown on a log$_{10}$ scale 
in Fig.1 at $\kappa=0.152$. 
It is the area under this curve which is of
interest here. The time integrals are very sharply peaked 
near $t=0$. The numerical quality of the vector time data is seen
to be quite good.

My {\it preliminary} result (at $\kappa=.154$) is (using the lattice
scale from Ref.\cite{scale})
$$
\alpha_{E} = 1.7 (2.1) (.8) \times 10^{-4} fm^{3}
$$
where the first number in parenthesis is the systematic uncertainty 
in the time integral, the second is statistical. 
The systematic uncertainty is defined to be half of the difference between
the discrete (Simpson's rule - smaller value) and continuous 
(free quark model - larger value) evaluations for $\alpha_{E}$. 
The above is to be compared with the experimental 
result (from MARK-II data\cite{mark2}),
$$
\alpha_{E}^{exp} = 2.2 (1.6) \times 10^{-4} fm^{3}
$$
and the result from second order chiral perturbation theory\cite{chiral},
$$
\alpha_{E}^{chiral} = 2.4 (.5) \times 10^{-4} fm^{3}.
$$
\begin{figure}
\vskip 65mm
\special{illustration 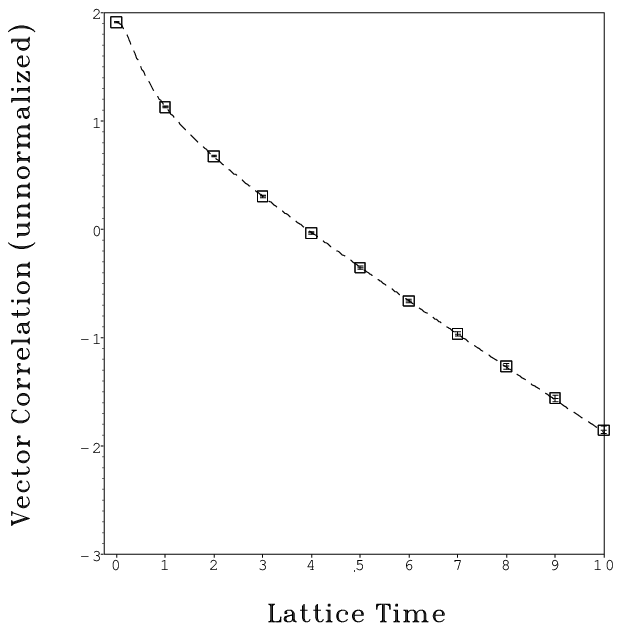}
\caption{Vector correlator at $\kappa =0.152$. The dashed line is the fit
from the free quark model.}
\label{figure2}
\end{figure}
So what one gets is surprisingly good, at least compared to experiment.
Warning: other sources of systematic error (momentum derivative, lattice scale) 
have not yet been evaluated.

\section{Summary and Remarks}

The sum rule method of extracting charged pion polarizability 
from lattice data has been examined. We have seen
that the systematic uncertainty in the time integral 
is larger than the statistical uncertainty, but still reasonable
compared to experiment.

To reduce this systematic uncertainty,
the situation here demands a fine mesh of lattice points in the time direction.
Coarse lattices are not useful in the case of such sum rules. 
However, time asymmetric lattices with a small time spacing
(but not too small to get into the asymptotic regime) could possibly be 
very useful.

Pion polarizability scales like $a^{3}$,
and so is extremely sensitive to the lattice scale. When the experimental
situation is improved, it could easily become one of the most
sensitive tests of QCD.

\section{Acknowledgements}

This work is supported in part by NSF Grant No 9401068
and the National Center for Supercomputing Applications. The author would
like to thank J.\ Vasut and B.\ Lepore for help with the free quark model integral.

\end{document}